# Quantitative measurements of size-dependent magnetoelectric coupling in $Fe_3O_4$ nanoparticles

Kyongjun Yoo[1], Byung-Gu Jeon[1], Sae Hwan Chun[1], Deepak Rajaram Patil[1], Yong-jun Lim[2,3,4], Seung-hyun Noh[2,3,4], Jihyo Gil[2,3,4], Jinwoo Cheon[*,2,3,4] and Kee Hoon Kim[*,1]

[1]*Center for Novel States of Complex Materials Research and Institute of Applied Physics, Department of Physics and Astronomy, Seoul National University, Seoul 151-747, S. Korea*

[2]*Center for Nanomedicine, Institute for Basic Science (IBS), Seoul 03722, S. Korea*

[3]*Yonsei-IBS Institute,Yonsei University, Seoul 03722, S. Korea*

[4]*Department of Chemistry, Yonsei University, Seoul 03722, S. Korea*

**Abstract**

Bulk magnetite ($Fe_3O_4$), the loadstone used in magnetic compasses, has been known to exhibit magnetoelectric (ME) properties below ~10 K; however, corresponding ME effects in $Fe_3O_4$ nanoparticles have been enigmatic. We investigate quantitatively the ME coupling of spherical $Fe_3O_4$ nanoparticles with uniform diameters ($d$) from 3 to 15 nm embedded in an insulating host, using a sensitive ME susceptometer. The intrinsic ME susceptibility (MES) of the $Fe_3O_4$ nanoparticles is measured, exhibiting a maximum value of ~0.6 ps/m at 5 K for $d$=15 nm. We found that the MES is reduced with reduced $d$ but remains finite until $d$=~5 nm, which is close to the critical thickness for observing the Verwey transition. Moreover, with reduced diameter, the critical temperature below which the MES becomes conspicuous increased systematically from 9.8 K in the bulk to 19.7 K in the nanoparticles with $d$=7 nm, reflecting the core-shell effect on the ME properties. These results point to a new pathway for investigating ME effect in various nanomaterials.

Magnetic nanoparticles have been extensively studied for the last decade due to their diverse biomedical applications, such as contrast agents for magnetic resonance imaging, hyperthermia, drug delivery, and bio separation.[1-4] However, their magnetoelectric (ME) properties, which allow for the modulation of electric polarization *P* (magnetization *M*) by a magnetic field *H* (electric field *E*)[5-6], have seldom been explored. The lack of work on ME properties mainly originates from the difficulties in determining the intrinsic ME coupling, particularly in the case of nanoparticles, due to their high electrical leakage and weak ME signals. Most previous studies on ME properties have thus been limited to bulk materials and thin composite films[7-9]. Meanwhile, ME nanoparticles, once realized, are expected to be useful for various applications, e.g., applying electrical stimuli for cell proliferation[10], preparing electrically responsive surface for functionalization[11], developing stimuli-responsive photonic crystals[12], and so on.

To search for possible ME nanoparticles, it is first necessary tounderstand the behavior of an archetypal ME material, namely the magnetite ($Fe_3O_4$), in nanoparticle form. As the oldest known loadstone material and with a ferrimagnetic Curie temperature of 858 K, $Fe_3O_4$ exhibits an intriguing metal-insulator transition, known as the Verwey transition, at $T_v$ =~123 K. The Verwey transition is accompanied by an intriguing charge ordering between $Fe^{2+}$ and $Fe^{3+}$ ions in the inverse spinel structure and a concomitant structural transition from a cubic to a monoclinic phase.[13-15] However, this unique but complex phase transition, involving entangled spin-charge-lattice degrees of freedom, is not fully understood, and remains one of the long-standing conundrums in material physics.[15] Apart from this Verwey transition, bulk $Fe_3O_4$ has been known to become ME below its ferroelectric transition, located at around 10-38 K[16-18], at which point a triclinic phase transition also seems to occur.[19] Rado *et al.*[7-8] were the first to observed the linear and bilinear ME effects in bulk $Fe_3O_4$ at 4.2 K. Several subsequent studies on the ME effect[20], heat capacity[21] and magnetization[22] have uncovered an anomaly in the vicinity of 10 K, below which the ME coupling becomes sufficiently sizable to be measured. There have also been theoretical predictions that the ME effect arises from the peculiar charge/orbital ordering pattern involved in the Verwey transition and from the *p-d* hybridization mechanism related to the intrinsic spin-orbit interaction.[23,24]

On the other hand, mainly due to the experimental difficulties described above, there has been no systematic study of the size dependence of ME properties in spherical $Fe_3O_4$ nanoparticles to date. However, progress in synthesis techniques has produced uniform, diameter-controlled nanoparticles, allowing for quantitative investigations of diameter-dependent physical properties.[25-28] For instance, a recent report[27] by Lee *et al*. showed that the Verwey transition temperature starts to decrease for $d \leq 20$ nm and eventually disappears for $d \leq 6$ nm.

Here, we report the ME susceptibility (MES) of spherical $Fe_3O_4$ nanoparticles with a controlled, uniform diameter at various temperature and magnetic field conditions. In order to overcome the leakage currents arising from the conductivity of the nanoparticles, we mixed them into an insulating host material (Stycast$^{TM}$ 1266). Furthermore, we employed a custom-designed ME susceptometer with a high charge oscillation sensitivity of $10^{-17}$C.[29-31] The maximum MES ($dP/dH$) of 0.6 ps/m could be measured for nanoparticles with $d$=15 nm at 5 K. The MES is generally reduced as $d$ is decreased, but remains finite for $d \geq \sim 5$ nm. Moreover the temperature at which MES starts to appear is increased in the smaller nanoparticles, which could be explained by the core-shell effect.

$Fe_3O_4$ nanoparticles with an averaged diameter ($d$=3, 5, 7, 12, and 15 nm) were prepared through one-pot thermal decomposition method.[27-28] Iron chloride ($FeCl_2$) and iron-2,4-pentadionate ($[Fe(acac)_3]$) were used as precursors, and were decomposed in the presence of the surfactants, oleic acid and oleylamine. The diameters and shapes of the nanoparticles were precisely controlled by varying the ratio of the metal precursors to the surfactants. In order to study the intrinsic ME coupling of the nanoparticles, the ligand, consisting of the surfactants, was removed by the ligand exchange method, which makes use of the tetramethylammonium hydroxide (TMAOH) and butyl alcohol (BuOH). Detailed procedures for the synthesis of the nanoparticles are given in the Supporting Information. For a comparative study, we also investigated two commercially available $Fe_3O_4$ bulk powders with different impurity levels; one (bulk #1) is from Sigma Aldrich (310069-Iron(II,III)oxide powder, <5 μm, 95 %) with average particle diameter of $d=3\pm1$ μm, and the other (bulk #2) is from Alfa Aesar (12962-Iron(II,III)oxide powder, 99.997 %) with particle diameters distributed from ~10 to 50 μm, denoted as $d > 10$ μm.

For MES and dielectric constant measurements, the $Fe_3O_4$ nanoparticles and bulk powders were mixed with an insulating host matrix (Stycast$^{TM}$ 1266) and solidified under $H= 0.4$ T in order to align them along the magnetic easy axis (Fig. 1a). The volume fractions of the bulk powder and the nanoparticles in the insulating matrices were estimated from the measured density of the solidified Stycast$^{TM}$ with and without guest particles, and were later used for normalization of the obtained MES data. The solidified specimens were cut into thin plate plates with surface parallel to the initial $H$ direction, to align the nanoparticles. Silver electrodes were painted on both sides of the plate, in order to make parallel capacitors. To investigate ME properties, a highly sensitive ME susceptometer was developed, in which a pair of solenoid coils were used to generate an AC magnetic field of $\mu_0 H_{ac} \sim 0.4$ mT at a frequency of 171 Hz (Fig. 1b). The modulated polarization in proportional to the applied AC magnetic field was measured using a high-impedance charge amplifier with a gain factor of $10^{12}$ V/C and a lock-in amplifier. Before the MES measurements, each sample was electrically poled at $E_p$ = 3.33 kV/cm at a temperature of 50 K and $H$ = 2 T, and then cooled to the target temperature. The electric field was subsequently turned off, followed by an electrical shorting of the electrodes. Since the MES measurement employs an AC technique, there exist in- and out-of-phase components. The in-phase signal (i.e., d$P$/d$H$) represents the intrinsic ME coupling, whereas the out-of-phase component can arise from electric leakage of the sample and/or the eddy currents mainly coming from the electrodes. We could verify that the extrinsic out-of-phase component was significantly reduced by embedding the nanoparticles in Stycast$^{TM}$ (Figure S6, in the Supplementary Information). The dielectric properties were measured using a capacitance bridge (AH 2550A) at a frequency of 1 kHz, and magnetization by a vibrating sample magnetometer. All the measurements were carried out inside PPMS$^{TM}$ (Quantum Design) as a function of $H$ (0 – 2 T) and $T$ (5 K – 50 K). Both AC and DC magnetic fields were applied perpendicular to the sample surface.

The diameter and shape of nanoparticles were investigated using transmission electron microscopy (TEM, JEM-2100). Figure 1c shows a TEM image with an averaged diameter of $d$=15 nm. The TEM images and the histogram of the measured $d$ of the $Fe_3O_4$ nanoparticles (Figure S1, supporting information) demonstrate that the $d$ distribution is quite narrow ($\sigma <$ 13 %) and the spherical shapes is uniform for all of the particles investigated. In addition, the X-ray diffraction data (Figure S2, Supporting Information), indicates the high purity of the $Fe_3O_4$ nanoparticles due to the absence of any impurity peaks. All of the diffraction peaks can be indexed with the cubic $Fe_3O_4$ spinel structure.

Figure 2a shows the variation of MES ($\equiv$d$P$/d$H$) as a function of the bias magnetic field $H_{dc}$ for $Fe_3O_4$ nanoparticles with $d$=15 nm at $T$=5 K. When a positive electric field poling (+$E_P$) is applied and the magnetic field is reduced from 2 T, the MES increases up to the maximum of ~0.6 ps/m at 0.09 T and then decreases almost linearly to cross the intercepts of

d*P*/d*H* (0.41 ps/m) and $H_{dc}$ (-0.06 T) axes. The MES behavior can be generally understood from a free energy $F(E, H)$ expansion under the application of electric field and magnetic fields,

$$F(E,H) = -PE - \mu_0 MH + \frac{1}{2}\varepsilon_0 \varepsilon E^2 + \frac{1}{2}\mu_0 \mu H^2 + \alpha EH + \frac{1}{2}\beta EH^2 + \gamma E^2 H + \frac{1}{2}\pi E^2 H^2 + \cdots \quad (1)$$

where $\varepsilon$ and $\mu$ ($\varepsilon_0$ and $\mu_0$) are the dielectric permittivity and magnetic permeability of the material (vacuum) and $\alpha$, $\beta$, $\gamma$, and $\pi$ are linear ($\alpha$), quadratic ($\beta$) and higher order ME coefficients. By minimizing $F$ with respect to $E$, the modulation of $P$ and d*P*/d*H* at $E = 0$ under low magnetic field regions can be described as

$$\Delta P(E,H) = P(E,H) - P(E,0) = \alpha H + \frac{1}{2}\beta H^2 + (\gamma H + \pi H^2)E + \cdots \quad (2)$$

$$\frac{dP}{dH}(0,H) = \alpha + \beta H + \cdots \quad (3)$$

Above equations show that $\Delta P(0, H)$ in low field regions can be approximately described by the *H*- linear and quadratic terms, if both $\alpha$ and $\beta$ are finite values. Note that, in the expression for d*P*/d*H*, the linear ME coefficient $\alpha$, if allowed in a material, remains as a constant, and the quadratic term $\beta$ is linear for a low *H* bias.

The MES in figure 2a shows linear behavior for a low *H* bias which is often observed in a medium with a dominant strain-induced quadratic ME effect.[29] However, the shape of MES is not exactly asymmetric with respect to the *H* sign reversal, as expected only in the quadratic ME media; the absolute value of the maximum in d*P*/d*H* is clearly larger than that of the minimum in *dP/dH*, indicating that a linear ME coupling effect ($\alpha$) should also exist. Another noticeable feature in Figure 2a is that the sign of d*P*/d*H* is reversed with the change of the electric poling from a positive ($+E_P$) to a negative ($-E_P$) value. This is an expected behavior in ferroelectric media.

Figures 2b and 2c show the $H$-dependence of dielectric constant $\varepsilon$ and the square of the magnetization $M^2$ at 5 K for $Fe_3O_4$ nanoparticles with $d$=15 nm. The $\varepsilon$ data forms a hysteresis loop with butterfly shape, exhibiting maximum values at $H \approx \pm 0.06$ T. Similar butterfly loops have been observed in materials such as $Mn_3O_4$[32], $NiCr_2O_4$[33], and $BiMnO_3$[34], in which the magnetodielectric effects satisfying the relation $\varepsilon \propto M^2$ exist due to the presence of spin-phonon coupling. It is noteworthy that the hysteresis appearing in the $M^2$-$H$ curve is much smaller than that of $\varepsilon$-$H$ curve. Furthermore, to determine whether the quadratic ME coupling, i.e., $P \propto M^2$, is sufficient to explain the ME coupling, we carefully compare the $dP/dH$ and $dM^2/dH$ curves.[35] The $dP/dH$ - $H$ curve is indeed markedly different from the $dM^2/dH$ - $H$ curve in Fig. 2d; in particular, it shows a less perfectly asymmetric line shape with respect to $H$-direction change (Figure 2a). This observation also supports that the idea that $Fe_3O_4$ nanoparticles with $d$=15 nm should have both linear and quadratic effects as known for the bulk magnetite. The polycrystalline $Fe_3O_4$ pellet studied here (bulk #1) also exhibited almost same line shape as that of $Fe_3O_4$ nanoparticles, further supporting the presence of both linear and quadratic ME effect (Figure S4, supporting information).

Figure 2e presents the MES data for $Fe_3O_4$ nanoparticles ($d$=15 nm) measured measured at 10, 15, 30 and 50 K. As temperature is increased, the MES value as well as the highest field at which MES disappears decrease rapidly, but the small MES signal with a maximum value of ~0.03 ps/m was persistently observable up to 50 K. As the MES at 50 K shows an almost perfectly asymmetric line shape with respect to the sign reversal of $H$, the ME coupling is likely to be purely quadratic. However, for the negative poling ($-E_P$) at 50 K, the MES signal did not exhibit a perfect sign reversal as compared with the $+E_P$ case. This implies that the specimen is not in a perfect ferroelectric state or a perfect magnetoelectric state at 50 K. The MES vs. $H$ curve at 30 K looks quite similar to that at 50 K. Therefore, the small quadratic-like ME signals observed at 30 and 50 K for the $+E_P$ case might come from the magnetostriction of the $Fe_3O_4$ nanoparticles and resultant length modulation of Stycast$^{TM}$ via interface strain coupling between the Stycast$^{TM}$ and the $Fe_3O_4$, which have small piezoelectric[36] and magnetostriction properties, respectively. On the other hand, we note that the d$P$/d$H$ curves below 15 K exhibit a sudden increase in the overall MES magnitude as well as a noticeable difference in the absolute values of the maximum and minimum value of d$P$/d$H$. Moreover, the MES shows the sign reversal with respect to the poling direction change as expected in ferroelectric or linear ME media. This represents clear evidence of the drastically enhanced linear ME effect that is inherent to the $Fe_3O_4$ nanoparticles ($d$=15 nm) below 15 K. Therefore, the MES data measured at low temperatures below 15 K should reflect the intrinsic ME coupling of the nanoparticles.

Since we have confirmed that the MES data below 15 K reflects the intrinsic ME properties of nanoparticles, we next investigate the MES evolution with various $d$ at 5 K, as summarized in Figure 3a. The maximum MES value for microparticles with $d$ >10 μm and $d$=3μm, and nanoparticles with $d$=15, 12, 7, and 5 nm, all of which were embedded in Stycast$^{TM}$, were found to be 0.44, 0.48, 0.6, 0.25, 0.16 and 0.03 ps/m, respectively. The data implies that the maximum MES signals are more or less similar for a diameter range of 15 nm < $d$ < ~10 μm. Most importantly, the MES maximum starts to decrease systematically from $d$=15 to $d$=7 nm before almost disappearing for $d$≤5 nm.

Based on the integration of the d$P$/d$H$ vs. $H$ data, we can extract $H$-dependent variation of the electric polarization ($\Delta P$-$H$) in the nanoparticles. Figure 3b summarizes the $\Delta P$-$H$ curves obtained for $+E_p$ in Fig. 3a. Consistent with the d$P$/d$H$ vs. $H$ behavior, both symmetric $\Delta P_{sym}$

($=(P(H)+P(-H))/2$) and asymmetric $\Delta P_{asym}$ ($=(P(H)-P(-H))/2$) polarizations coexist in the Fe$_3$O$_4$ nanoparticles, similar to the case of the Fe$_3$O$_4$ single crystal.[8] $\Delta P_{sym}$ and $\Delta P_{asym}$ indeed do correspond to the quadratic and linear ME effects, respectively, in the low $H$ bias regime. It is found that $\Delta P_{sym}$ starts to decrease monotonically from $d=3$ μm, whereas $\Delta P_{asym}$ decreases from $d = 15$ nm. Eventually, both $\Delta P_{sym}$ and $\Delta P_{asym}$ are suppressed such that they become negligibly small in nanoparticles with $d = 5$ nm and completely disappear for d = 3nm. The critical diameter for having an appreciable ME effect is then at least 5 nm. Interestingly, Lee *et al.* recently found that the Verwey transition disappears in nanoparticles with d ≤ 6 nm, which is quite close to the critical diameter for observing the ME effect.[27] Therefore, our experimental findings corroborate the theoretical prediction that the ME effect in magnetite is linked to the peculiar charge-orbital-spin ordered state realized by the Verwey transition.

In sharp contrast to the decrease in the MES signal in nanoparticles with smaller diameters, we found that the MES signal survives for higher temperatures. Figure 4a presents the MES of the nanoparticles as a function of temperature from 5 to 50 K, measured at the magnetic field where MES shows a maximum at 5 K. As clearly seen in Fig. 4a, the temperature $T_{ME}$ below which the MES starts to be significant, has systematically increased with receded $d$ for both $+E_p$ and $-E_p$ cases. To estimate $T_{ME}$ consistently for each nanoparticle, we chose the temperature at which a base line and a linearly extrapolated line from the steeply increasing part of the d$P$/d$H$ curve meet. The dashed lines in Fig. 4a summarize $T_{ME}$'s values; 9.8, 15.6, 18.5, and 19.7 K for the micro-particles (bulk #2) with $d= 3$ μm, and nanoparticles with $d=$ 15, 12, and 7 nm, respectively [the corresponding curves for bulk #1 with $d>10$ μm were almost same as that with $d=3$ μm].

To understand the physical origin of the increase in $T_{ME}$ with reduced $d$, we measured $M$ vs

$T$ curves at 10 mT and plotted their temperature derivative d$M$/d$T$ in Figure 4b. An anomalous peak (or maximum) in the derivative of magnetization (d$M$/d$T$) appears at 9.9 K for the Fe$_3$O$_4$ microparticles, indicating that the magnetization sharply decreases at the temperature. It turns out that the peak temperature is indeed close to the $T_{ME}$ = 9.8 K, below which the triclinic structural transition presumably occurs, increasing the ME effects in bulk Fe$_3$O$_4$. Upon decreasing the particle diameter, the temperature of the d$M$/d$T$ maximum increases monotonically in rough proportion to $T_{ME}$; microparticles, the d$M$/d$T$ maximum almost coincides with $T_{ME}$, while for nanoparticles, it becomes slightly smaller overall. A close look at Fig. 4c reveals that the d$M$/d$T$ curves become progressively broader and exhibit weaker anomalies as the particle size is reduced. This indicates that the drop in magnetization gets shallower and broader. Therefore, the magnetization drop seems to occur in a broader temperature window as the surface effects become more important with the reduction of the particle diameter.

One appealing scenario to explain the above findings is the "core-shell" model expected in magnetic nanoparticles. In a recent study of related spinel CoFe$_2$O$_4$ nanoparticles which display a ferrimagnetic ordering at 860K in their bulk state, Sun *et al*[37] found spectroscopic (phonon vibration modes) evidence that a core with aligned spins is surrounded by magnetically disordered shell. Moreover, the vibrational mode analyses showed that the shell thickness of 0.4 nm in $d$=14 nm particles increased to 0.8 nm in 5 nm particles, demonstrating that an associated local distortion of the lattice takes place on the length scale of the unit cell. These experimental results by Sun *et al.* can be directly applied to understanding the size-dependent ME effect as well as the magnetization drop in the Fe$_3$O$_4$ nanoparticles. As the particle diameter is reduced, the magnetically disordered shell should increase in thickness. Moreover, the magnetically disordered shell seems to favor the stabilization of the triclinic structure over the monoclinic one at higher temperatures, possibly because the lower crystal symmetry becomes more compatible with the magnetically disordered spin state. As a result, the shell regions can display a concomitant structural and magnetic transition at high temperatures, the value which will have a distribution depending on the distance of each structural unit from the surface. The smaller the particle diameter, the broader can be the range of transition temperatures in the shell. Combined with the original magnetic transition in the core, the effective center temperature for the magnetic transition can be shifted toward higher temperatures, as illustrated in Fig. 4 (d). The conspicuously enhanced $T_{ME}$ in nanoparticles can then be attributed to the increased shell volume, which favors the stabilization of the magnetically disordered spin state and the triclinic phase at higher temperatures. We note that the present mechanism of the core-shell effect on ME properties is generally applicable to other nanomagnets exhibiting the ME effects.

In summary, we successfully measured the ME coupling of spherical $Fe_3O_4$ nanoparticles with systematic variation of the diameter from $d$=15 nm to 3 nm, by embedding the particles into a polymer host. When the diameter of $Fe_3O_4$ nanoparticles is reduced, at the lowest temperature of 5 K, the ME coupling becomes weaker and eventually disappears at a critical diameter of 5 nm. On the other hand, the ME coupling persists to higher temperatures up to ~ 20 K in the nanoparticles with $d$=7 nm, which could be understood as a result of the shell effect with the magnetically disordered spin state. The observation of size-dependent ME properties in $Fe_3O_4$ nanoparticles should further stimulate the exploration of new ME effects and their applications in other nanomaterials.

**Associated Content**

Supporting information

Experimental details, TEM images and size distribution histograms, magnetic field dependence of dielectric constant curve, MES data of $Fe_3O_4$ powder without Stycast$^{TM}$, zoomed up temperature dependent MES in particles with various diameters, in- and out-of-phase of MES data of $Fe_3O_4$ particles with and without Stycast$^{TM}$

**Author Information**

Corresponding Authors

*Email: khkim@phya.snu.ac.kr.

*Email: jcheon@yonsei.ac.kr.

Notes

The authors declare no competing financial interest.

**Acknowledgments**

This work was financially supported by National Creative Research Initiative (2010-0018300), Korea-Taiwan Cooperation Program (0409-20150111), and Global R&D Center (2016K1A4A3914691), and Institute for Basic Science (IBS-R026-D1) through the NRF of Korea funded by the Ministry of Science, ICT and Future Planning.

**References**

1. Laurent, S.; Forge, D.; Port, M.; Roch, A.; Robic, C.; Vander Elst, L.; Muller, R. N. *Chem. Rev.* **2008**, *108*, 2064-2110.

2. Wu, W.; He, Q.; Jiang C. *Nanoscale Res. Lett.* **2008**, *3*, 397-415.

3. Jun, Y.-W.; Seo, J.-W.; Cheon, J. *Acc. Chem. Res.* **2008**, *41*, 179-189.

4. Lee, N.; Hyeon, T. *Chem. Soc. Rev*. **2012**, *41*, 2575-2589.

5. Kimura, T.; Goto, T.; Shintani, H.; Ishizaka, K.; Arima, T.; Tokura, Y. *Nature* **2003**, *426*, 55-58.

6. Hur, N.; Park, S.; Sharma, A.; Ahn, J. S.; Guha, S.; Cheong, S. W. *Nature* **2004**, *429*, 392-395.

7. Rado, G. T.; Ferrari, J. M. *Phys. Rev. B* **1975**, *12*, 5166-5174.

8. Rado, G. T.; Ferrari, J. M. *Phys. Rev. B* **1977**, *15*, 290-297.

9. Zheng, H.; Wang. J.; Lofland S. E.; Ma, Z.; Mohaddes-Ardabili, L.; Zhao, T.; Salamanca-Riba, L.; Shinde, S. R.; Ogale, S. B.; Bai, F.; Viehland, D.; Jia, Y.; Schlom, D.G.; Wuttig, M.; Roytburd, A.; Ramesh, R. *Science* **2004**. *303*, 661-663.

10. Ribeiro, C.; Correia, V.; Martins, P.; Gama, F. M.; Lanceros-Mendez, S. *Colloids Surf. B* **2016,** *140,* 430-436.

11. Cantini, E.; Wang, X.; Koelsch, P.; Preece, J. A.; Ma, J.; Mendes, P. M. *Acc. Chem. Res.* **2016,** *49,* 1223-1231.

12. Lu, T.; Peng, W.; Zhu, S.; Zhang, D. *Nanotechnology* **2016,** *27,* 122001.

13. Verwey, E. J. *Nature* **1939**, *144*, 327-328.

14. Walz, F. *J. Phys.: Condens. Matter* **2002**, *14*, R285-R340.

15. Senn, M. S.; Wright, J. P.; Attfield, J. P. *Nature* **2012**, *481*, 173-176.

16. Miyamoto, Y.; Kobayashi, M.; Chikazumi, S. *J. Phys. Soc. Japan* **1986**, *55*, 660-665.

17. Alexe, M.; Ziese, M.; Hesse, D.; Esquinazi, P; Yamauchi, K.; Fukushima, T.; Picozzi, S.; Gosele, U. *Adv. Mater.* **2009**, *21*, 4452-4455.

18. Ziese, M.; Esquinazi, P. D.; Pantel, D.; Alexe, M.; Nemes, N. M.; Garcia-Hernandez, M. *J. Phys.: Condens. Matter* **2012**, *24*, 086007.

19. Medrano, C.; Schlenker, M.; Baruchel, J.; Espeso, J.; Miyamoto, Y. *Phys. Rev. B* **1999**, *59*, 1185-1195.

20. Miyamoto, Y.; Ariga, M.; Otuka, A.; Morita, E.; Chikazumi, S. *J. Phys. Soc. Japan* **1979**, *46*, 1947-1948.

21. Todo, S.; Chikazumi. S. *J. Phys. Soc. Japan* **1977**, *43*, 1091-1092.

22. Matsui, M.; Todo, S.; Chikazumi. S. *J. Phys. Soc. Japan* **1977**, *43*, 47-52.

23. Brink, J.; Khomskii, D. I. *J. Phys.: Condens. Matter* **2008**, *20*, 434217.

24. Yamauchi, K.; Picozzi, S. *Phys. Rev. B* **2012**, *85*, 085131.

25. Sun, S.; Zeng, H.; Robinson, D. B.; Raoux, S.; Rice, P. M.; Wang, S. X.; Li, G. *J. Am. Chem. Soc.* **2003**, *126*, 273-279.

26. Park, J.; An, K.; Hwang, Y.; Park, J.-G.; Noh, H.-J.; Kim, J.-Y; Park, J.-H.; Hwang, N.-M.; Hyeon, T. *Nat. Mater.* **2004**, *3*, 891-895.

27. Lee, J.; Kwon, S. G.; Park, J.-G; Hyeon, T. *Nano Lett.* **2015**, *15*, 4337-4342.

28. Noh, S.–H.; Na, W.; Jang, J.-T.; Lee, J.-H.; Lee, E. J.; Moon, S. H.; Lim, Y.; Shin, J.-S.; Cheon, J. *Nano Lett*. **2012**, *12*, 3716-3721.

29. Oh, Y. S.; Crane, S.; Zheng, H.; Chu, Y. H.; Ramesh, R.; Kim, K. H. *Appl. Phys. Lett.* **2010**, *97*, 052902.

30. Ryu, H.; Murugavel, J. H.; Lee. J. H.; Chae, S. C.; Noh, T. W.; Oh, Y. S.; Kim, H. J.; Kim, K. H.; Jang, J. H.; Kim, M.; Bae, C.; Park, J.G. *Appl. Phys. Lett.* **2005**, *89*, 102907.

31. Ko, K. –T.; Jung, M. H.; He, Q.; Lee, J. H.; Woo, C. S.; Chu, K.; Seidel, J.; Jeon, B. –G.; Oh, Y. S.; Kim, K. H.; Liang, W. –I.; Chen, H. –J.; Chu, Y. –H.; Jeong, Y. H.; Ramesh, R.; Park, J. –H.; Yang, C. –H. *Nature Commun.* **2011,** *2,* 567.

32. Tackett, R.; Lawes, G.; Melot, B. C.; Grossman, M.; Toberer, E. S.; Seshadri, R. *Phys. Rev. B* **2007**, *76*, 204409.

33. Sparks, T. D.; Kemei, M. C.; Barton, P. T.; Seshadri, R.; Mun, E.-D.; Zapf, S. *Phys. Rev. B* **2014**, *89*, 204405.

34. Kimura, T.; Kawamoto, S.; Yamada, I.; Azuma, M.; Takao, M.; Tokura, Y. *Phys. Rev. B* **2003**, *67*, 180401.

35. Zhou, Y.; Yang, S. C.; Apo, D. J.; Maurya, D.; Priya, S. *Appl. Phys. Lett.* **2012**, *101*, 232905.

36. Troge, A.; O'Leary, R. L.; Hayward, G.; Pethrick, R. A.; Mullholland A. J. *J. Acoust. Soc. Am* **2010**, *128*, 2704.

37. Sun, Q. –C.; Birkel, C. S.; Cao, J. B.; Tremel, W.; Musfeldt, J. L.; *ACS nano* **2012,** *6,* 4876.

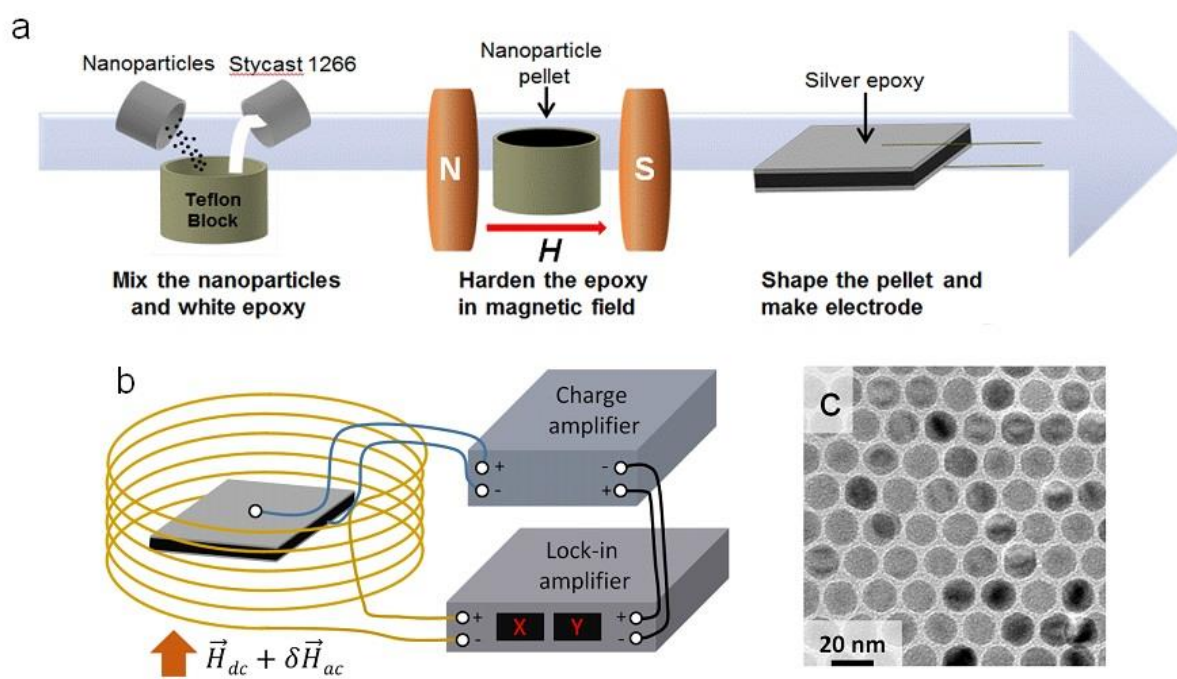

**Figure 1** (a) Schematic illustration of the preparation of the nanoparticles and (b) the configuration of the ME measurements employed in the magnetoelectric susceptometer. (c) Transmission electron microscopy images of $Fe_3O_4$ nanoparticles with a diameter of 15 nm.

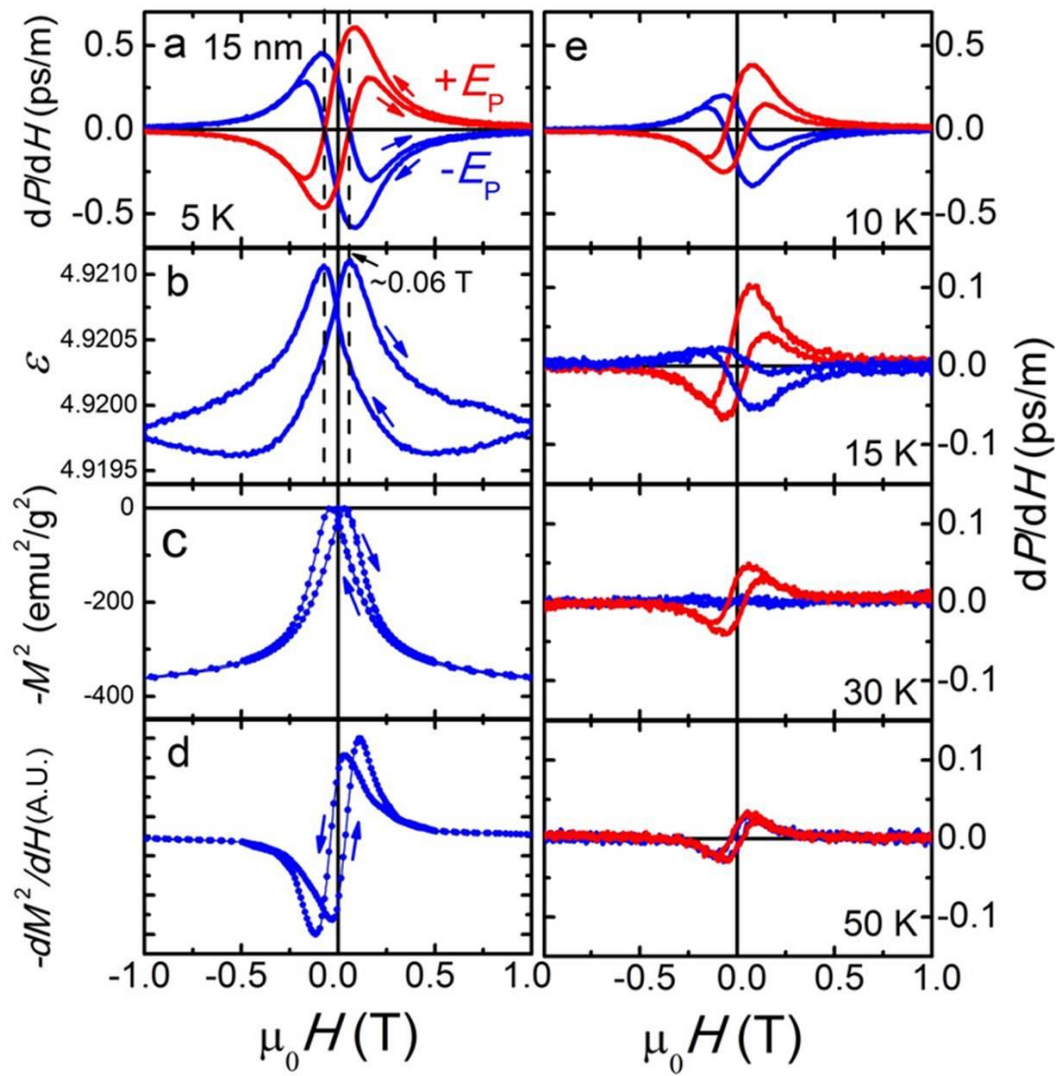

**Figure 2** Magnetic field dependence of (a) magnetoelectric susceptibility (MES) ($=dP/dH$), (b) dielectric constant ($\varepsilon$), (c) square of magnetization at 5 K, (d) field derivative of square of magnetization at 5 K and (e) MES of the $Fe_3O_4$ nanoparticles with $d=$ 15 nm at selected

temperatures.

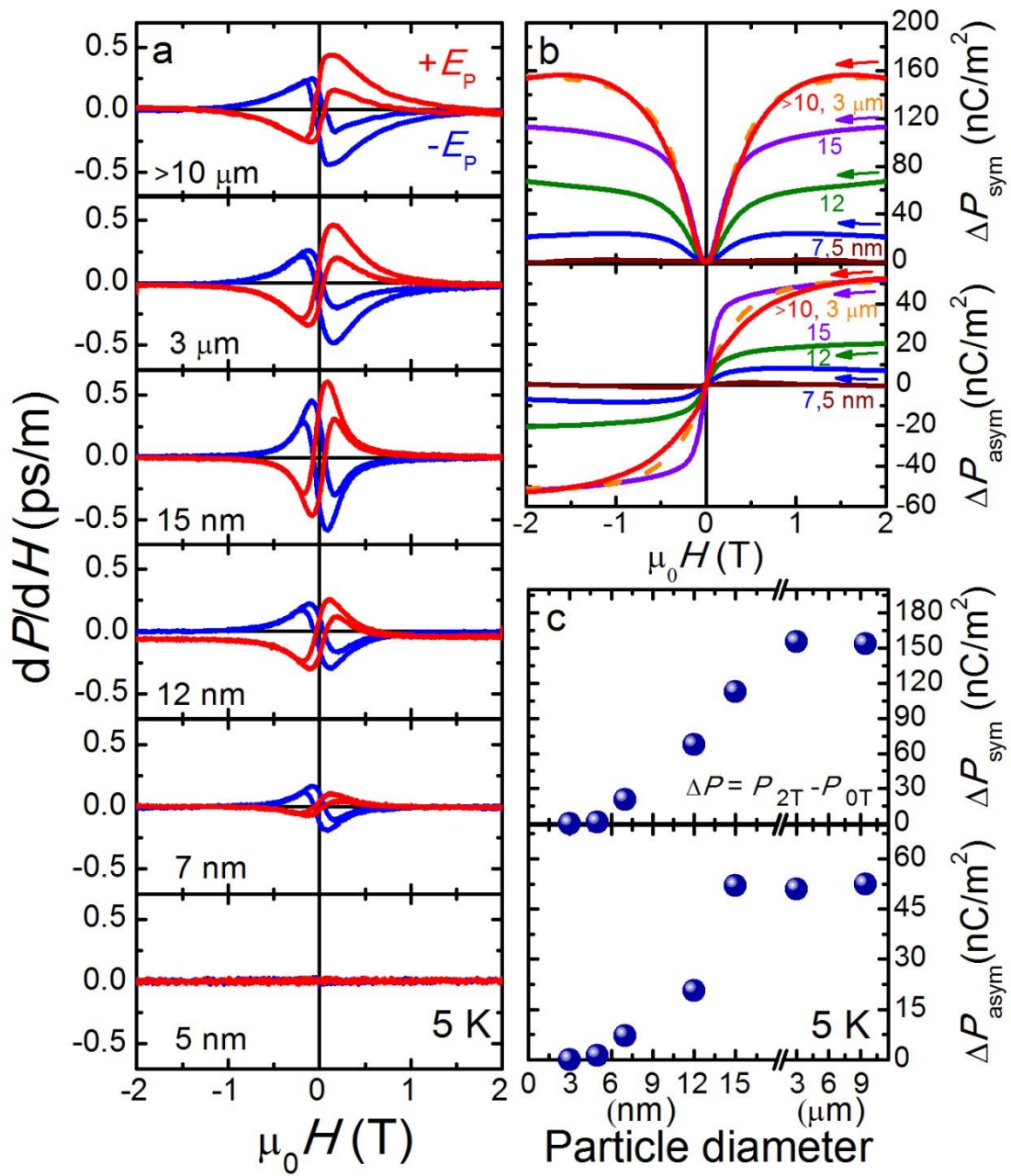

**Figure 3** Magnetic field dependence of (a) d$P$/d$H$ and (b) polarization integrated from d$P$/d$H$ cureve resulting in symmetric and asymmetric shapes with respect to $H$ in the $Fe_3O_4$ particles of various diameterss. (c) The polarization difference between 0 and 2 T for the symmetric and asymmetric cases.

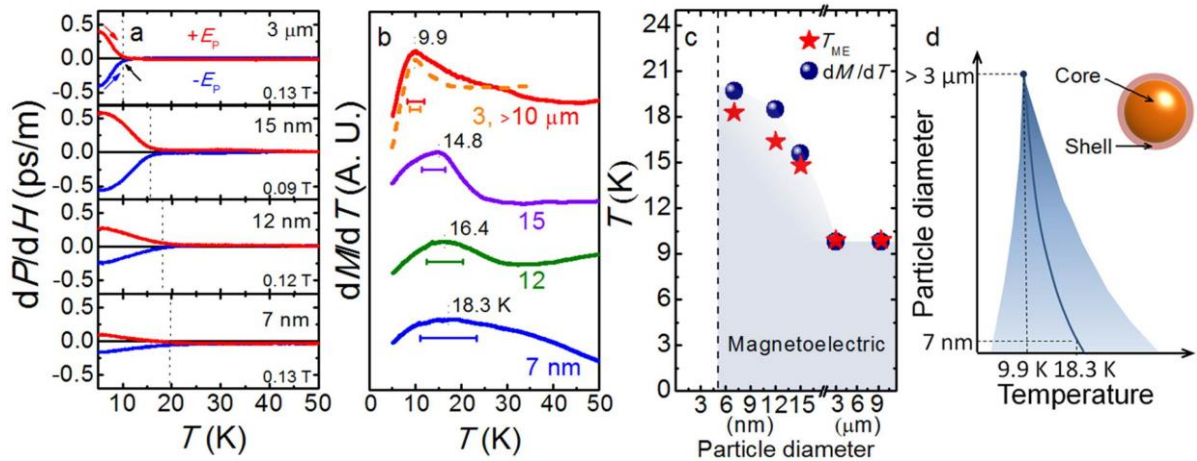

**Figure 4** (a) Temperature dependence of d$P$/d$H$ for $Fe_3O_4$ particles with $d$=3 μm, 15 nm, 12 nm, and 7 nm, all embedded in the Stycast$^{TM}$. (b) Temperature derivative of the magnetization (d$M$/d$T$) measured at $H$= 10 mT. The horizontal bar in each plot represents a temperature window that shows 3 % variation from the maximum value. (c) The maxima from the d$M$/d$T$ curves (asterisks) and $T_{ME}$ (solid circles) where the ME effect starts to increase steeply. (d) Illustration of the core-shell effect on the ME effect, resulting in the effective increase of $T_{ME}$ and the broadening of the d$M$/d$T$ curves in smaller particles.